\documentclass[amssymb,nofootinbib,10.5 pt]{revtex4}
\usepackage{graphicx}
\begin{document}
\title{A classification of entanglement in three-qubit systems}
\author{Carlos Sab\'in}\affiliation{Instituto de Matem\'aticas y F\'isica Fundamental,CSIC, Serrano, 113-bis, 28006 Madrid (Spain)}
\email{csl@imaff.cfmac.csic.es}\author{Guillermo Garc\'ia-Alcaine} \affiliation{Departamento de F\'isica Te\'orica I,
Universidad Complutense de Madrid, Ciudad Universitaria, 28040 Madrid (Spain)} \email{ggarciaa@fis.ucm.es}
%\date{}
\begin{abstract}
We present a classification of three-qubit states based in their three-qubit and reduced two-qubit entanglements. For pure
states these criteria can be easily implemented, and the different types can be related with sets of equivalence classes under
Local Unitary operations. For mixed states characterization of full tripartite entanglement is not yet solved in general; some
partial results will be presented here.
\end{abstract}
\maketitle
\section{Introduction}
\label{intro}

Entanglement is a fundamental property of quantum systems and a basic resource for Quantum Information and Computation. However,
its detection and quantification are solved only in some simple cases, like the one of two-qubit systems. Discriminating
separability from entanglement for higher dimensional non-pure two-particle states seems to be in the complexity class NP-HARD
\cite{gur}. Finding good measures of entanglement is a related nontrivial problem, even for the next simplest possibilities:
non-pure two-particle states in arbitrary dimensions or three-qubit non-pure states.

For three-qubit pure states some authors \cite{aciuno}, \cite{acidos} divide the infinite set of equivalence classes under Local
Unitary (LU) transformations in subsets characterized by particular values of some LU invariants: these invariants are not
directly related with entanglement measures, and therefore this is not exactly a classification of entanglement types. Other
authors \cite{ciruno}, \cite{luc} classify three-qubit states in equivalence classes under Stochastic Local Operations and
Classical Communication (SLOCC); some of the types considered in other classifications, like the \textit{star-shaped states}
introduced in \cite{pleuno}, \cite{pledos} do not have their own class in this scheme.

Several entanglement measures for pure states of three-particle systems have been proposed \cite{coff}, \cite{chanuno},
\cite{mey}, \cite{bren}, \cite{pan}, \cite{pas}; as we will discuss in \ref{sec:3}, these measures do not distinguish adequately
between fully entangled and separable states. Extensions to mixed states are even more problematic; we will discuss a measure
that improves on these results.

The paper is organized as follows. In \ref{sec:1} we summarize some known results for the entanglement of two and three-qubit
systems, including definitions and notation that we will use in the following sections. In \ref{sec:2} we introduce a
classification for arbitrary  three-qubit states in terms of their three-qubit and reduced two-qubit entanglements, with a
graphic representation of the different types for pure states. In \ref{sec:3} we comment several previous measures of tripartite
entanglement, and introduce a multiplicative generalization of two-particle \textit{negativity}. In \ref{sec:4} we apply a
Generalized Schmidt Decomposition (GSD) \cite{aciuno} to pure three-qubit states and specify the form of the states in each of
our classes; this could be used to give a physical interpretation to the abstract classes of \cite{aciuno}. In \ref{sec:5}, we
calculate the tripartite negativity of some families of mixed states and compare our results with previously published ones. We
conclude in \ref{sec:6} with a summary of our results.
\section{Entanglement in two and three-qubit systems}
\label{sec:1}

Arbitrary states of pure two-qubit systems $|{\Psi}\rangle=\\ \sum_{i,j}\alpha_{ij}|{ij}\rangle$ , $|{ij}\rangle$ $(i,j=0,1)$
being a basis of the Hilbert space $H_{A}\otimes H_{B}$ for qubits A and B, can always be converted in
$|{\Phi}\rangle=\alpha_{00}|{00}\rangle+\alpha_{11}|{11}\rangle$, by means of LU transformations $U=U_{A}\otimes U_{B}$, being
 $U_{A}$ and $U_{B}$ unitary operators acting on $H_{A}$ and $H_{B}$ respectively. Each pair ${\alpha_{00},\alpha_{11}}$ of this
Schmidt decomposition define a LU equivalence class; this infinite set can be divided in two big subsets: \textit{separable
states} if one of the  $\alpha_{00},\alpha_{11}$ is zero, and \textit{entangled states} otherwise; generalization to pure states
in arbitrary dimensions is easy. For pure states, separability equals factorizability.

For the non-pure case, a state $\rho$ is separable \cite{wer} if it can be written as
$\rho=\sum_{i}p_{i}\rho_{i}^{A}\otimes\rho_{i}^{B}$, where $\rho_{i}^{A}$ and $\rho_{i}^{B}$ are state operators of subsystems A
and B respectively, $p_{i}\geq0 $ $\forall i$, $\sum_{i}p_{i}=1$; otherwise $\rho$ is entangled. A list of criteria for
separability in two-particle systems can be found in \cite{rad}. Sometimes \cite{wer}, \cite{pledos}, separable but not
factorizable mixed states are qualified as \textit{classically correlated}; we will not use this distinction, because we are
interested only in quantum correlated states.

We shall cite only three of the several entanglement measures of pure two-qubit systems: \textit{von Neumann's entropy}
\cite{sak} of reduced states, \textit{Wootters concurrence} \cite{woo}, and \textit{negativity} \cite{vidwer}.

The Von Neumann's entropy of a state $\rho$ is defined in Information Theory as $S(\rho)=-\sum_{j}p_{j}\log_{2}p_{j}$, $p_{j}$
being the eigenvalues of $\rho$.The reduced states are $\rho^{(A)}=Tr_B\, \rho$ and $\rho^{(B)}=Tr_A\, \rho$. Von Neumann's
entropy of reduced states is the simplest measure of two-particle entanglement for pure states, but its extension to non-pure
states discriminates between separable and entangled states only if the \textit{mutual entropy} or \textit{correlation index} is
zero   \cite{bar}. Therefore, it is not a good measure of entanglement for general mixed states.

Concurrence is defined as $C(\rho)=\\ max{\{0,\sqrt{\lambda_{1}}-\sqrt{\lambda_{2}}-\sqrt{\lambda_{3}}-\sqrt{\lambda_{4}}}\},$
where ${\{\lambda_{i}}\}$ $(i=1...4)$ are the eigenvalues of $R=\rho\widetilde{\rho}$  in decreasing order, with
$\widetilde{\rho}=(\sigma_{y}\otimes\sigma_{y})\rho^{\star}(\sigma_{y}\otimes\sigma_{y})$, $\rho^{\star}$ being the complex
conjugate of the state operator $\rho$. In arbitrary dimensions, simple computable generalizations of concurrence are known only
for pure states \cite{rung}; there is no efficient way known \cite{rew} to calculate sophisticate generalizations like
biconcurrence \cite{bad}.

The negativity of a bipartite state $\rho$ was introduced in \cite{vidwer}; we will use the convention of \cite{mir}, that is
twice the value of the original definition:
\begin{equation}
N(\rho)=-2\sum_{i}\sigma_{i}(\rho^{TA}), \label{eq:1}
\end{equation}
where ${\{\sigma_{i}(\rho^{TA})}\}$ are the negative eigenvalues of the partial transpose $\rho^{TA}$ of the total state $\rho$
respect to the subsystem A, defined as
$\langle{i_{A},j_{B}}|{\rho^{TA}}|{k_{A},l_{B}}\rangle=\langle{k_{A},j_{B}}|{\rho}|{i_{A},l_{B}}\rangle$, A and B denoting the
two subsystems. For pure bipartite states of arbitrary dimensions the negativity (1) is equal to the concurrence \cite{mirdos}.
The negativity is not additive \cite{plen} and some authors prefer to use the logarithmic negativity, which is additive but not
convex \cite{plendos}. The negativity can be evaluated in the same way for pure and non-pure states in arbitrary dimensions,
although there are entangled mixed states with zero negativity in every dimensions except $2\times2$ and $2\times 3$
\cite{horuno}, \cite{hordos}. No measure discriminating separable from entangled states in the general non-pure case is known.
However, non-zero negativity is a sufficient condition for entanglement.

For three qubits, there is no ternary Schmidt decomposition of pure states in a strict sense, but there is a Generalized Schmidt
Decomposition (GSD) \cite{aciuno}: arbitrary states can always be converted to states that contain at most five of the eight
$\alpha_{ijk},(i,j,k=0,1)$ terms of the Hilbert basis, by means of a very simple algorithm. Each set of coefficients
${\alpha_{000},\alpha_{100}, \\ \alpha_{101},\alpha_{110},\alpha_{111}}$ define a LU equivalence class; in \ref{sec:2} and
\ref{sec:4} we deal with the problem of dividing the infinite set of LU equivalence classes in big subsets corresponding to the
different types of entanglement in our classification.

For three-particle systems, a pure state $|{\Psi}\rangle$ is \textit{fully separable} if it can be written as
$|{\Psi}\rangle=|{\phi}\rangle_{A}\otimes|{\psi}\rangle_{B}\otimes|{\chi}\rangle_{C}$, \textit{biseparable} if it is not fully
separable but can be written as $|{\Psi}\rangle=|{\phi}\rangle_{A}\otimes|{\Phi}\rangle_{BC}$ with $|{\Phi}\rangle_{BC}$
entangled (the index A can denote any of the three subsystems), and \textit{fully inseparable} otherwise; in this last case we
will say that the state has \textit{full tripartite entanglement}.

For three-particle systems a non-pure state operator $\rho$ may be fully separable, biseparable or fully inseparable;
biseparable states can be \textit{simply biseparable} or \textit{generalized biseparable}. Fully separable states can be written
as $\rho=\sum_{i}p_{i}\ \rho_{i}^{A}\otimes\rho_{i}^{B}\otimes\rho_{i}^{C}$. Biseparable states are not fully separable but can
be written as $\rho=\sum_{iJ}p_{iJ}\ \rho_{i}^{J}\otimes\rho_{i}^{KL}$ where $J$ runs from A to C and $KL$ from $BC$ to $AB$
respectively and at least one $\rho_{i}^{KL}$ is entangled \cite{acitres}, \cite{dru}; simply biseparable states have
$p_{iJ}\neq0$ only for a single value of $J$ (one single qubit is separable from the other two, that are entangled); generalized
biseparable states are convex sums of states of the previous kind (non vanishing coefficients $p_{iJ}$ for more than one $J$).
Fully inseparable states are those not fully separable nor biseparable. Not all authors agree with this classification of mixed
three-qubits: for some of them, biseparable states were only those that we call simply biseparable, and generalized biseparable
states were included in the fully inseparable class \cite{cirdos}, \cite{egg}.

Fully separable estates contain no quantum entanglement. Simply biseparable states have bipartite entanglement in a single pair
of qubits; they have \textit{partial bipartite entanglement}. Generalized biseparable states contain bipartite entanglement in
more than one pair of qubits; they have \textit{distributed bipartite entanglement}. Fully inseparable states have full
tripartite entanglement (true nonclassical 3-particle correlations, in the notation of \cite{lask}).

The characterization of separability or biseparability for non-pure three-particle states, and the measure of their full
tripartite entanglement are nontrivial, even in the simplest case of three qubits. An ideal measure of tripartite entanglement
would discriminate fully entangled from fully separable or biseparable states; no such measure is known yet, even for three
qubits.

Finally, entanglement of the reduced two-particle states will be called the \textit{reduced entanglement} of the pair. They show
how much two-qubit entanglement remains when the third qubit is not observed. As we will show in the next section, for non-pure
states there are unexpected results; for instance, some simply biseparable mixed states have no reduced entanglement when the
separable qubit is traced over.
\section{A classification of entanglement in three-qubit systems}
\label{sec:2}

We propose a classification of three-qubit states based in the presence of no entanglement (full separability), bipartite
entanglements only (simple or generalized biseparability), or full tripartite entanglement (true 3-qubit entanglement, full
inseparability), both for pure and non-pure states, with subtypes based on the number of entangled reduced two-qubit states.
Reduced, bipartite and full tripartite entanglements have direct physical meaning and are LU invariants: local operations cannot
create entanglement, and therefore all the invertible local operations (unitary ones in particular) must leave these
entanglements invariant. Thus, a state in any of our subtypes cannot be transformed by a LU to a state in any other subtype.

This classification is equally valid for pure and non-pure states, although there are several subtypes that exist only for
non-pure ones. The practical implementation of the classification is also more difficult for mixed states, since the detection
of tripartite entanglement is not completely solved in this case. Another difference is that for pure states we can relate these
subtypes with the coefficients of a GSD, as we will show in \ref{sec:4}, while no similar decomposition exists for non-pure
states.

Fig.1 summarizes the different types and subtypes in this classification for pure states; with the same conventions, we could
draw a similar graphic for the subtypes that exist only for mixed states; we omit it for simplicity.

\begin{figure}[t]
\begin{center}
  \includegraphics[width=0.9\textwidth]{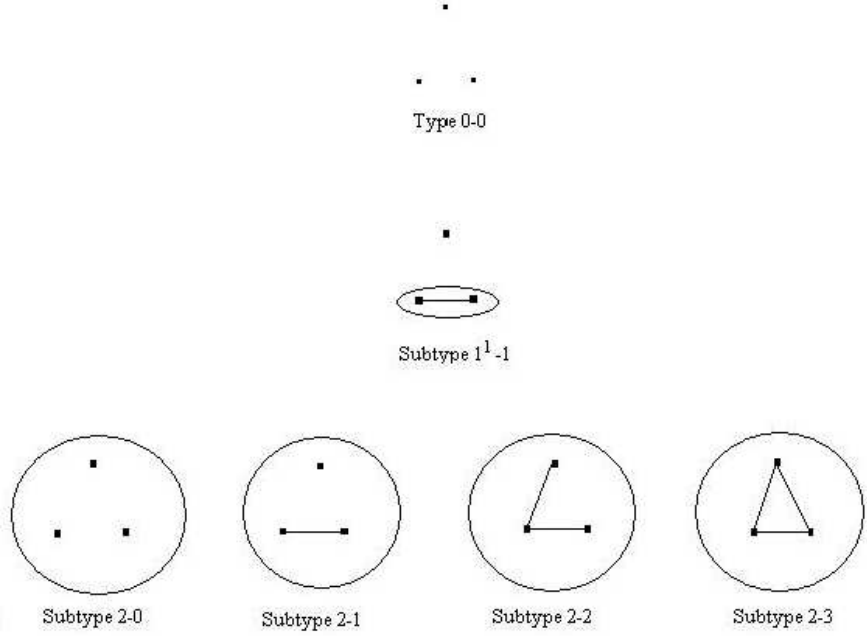}
\caption{A graphic representation of entanglement types in pure states of three-qubit systems. Each point represents one qubit.
A circle including the three qubits denotes full tripartite entanglement. An ellipse around two qubits means that they are
entangled, and the total three-qubit state is simply biseparable. A segment connecting two qubits means that they have reduced
two-qubit entanglement when the opposite qubit is traced over.}
\end{center}
\label{Fig:1}
\end{figure}
We will briefly comment now the different subtypes for general states (pure or not).
\begin{itemize}
\item Type $0-0$: fully separable states, no quantum entanglement.
\item Type 1: biseparable states, bipartite entanglements only.
\item Subtype $1^1-0$: simply biseparable states with no reduced entanglement. Two of the qubits are entangled when the total three-qubit
state is considered, but when the other qubit is traced over, the reduced two-particle state is separable. This subtype exists
only for non-pure states: for instance,  a biseparable state\\
$\rho={1\over2}|1\rangle\langle1|_{A}\otimes|\Psi_{+}\rangle\langle\Psi_{+}|_{BC}+{1\over2}|0\rangle\langle0|_{A}\otimes|\Psi_{-}\rangle
\langle\Psi_{-}|_{BC}$, where $|\Psi_{\pm}\rangle={1\over\sqrt{2}}(|10\rangle\pm|01\rangle)$ are Bell states, has a reduced
state $\rho^{(BC)}={1\over2}|\Psi_{+}\rangle\langle\Psi_{+}|_{BC}+{1\over2}|\Psi_{-}\rangle \langle\Psi_{-}|_{BC}$ that is
separable: $\rho^{(BC)}={1\over2}|1\rangle\langle1|_{B}\otimes|0\rangle\langle0|_{C}+{1\over2}
|0\rangle\langle0|_{B}\otimes|1\rangle\langle1|_{C}$; this is a counterintuitive result. The graphic representation of this
subtype would be the same as subtype $1^1-1$ (Fig.1) without the segment connecting the lower two qubits.
\item Subtype $1^1-1$: the three-qubit state is simply biseparable, and when the separable qubit is traced over, the other two remain
entangled. All pure biseparable states are of this subtype: any pure state
$|\Psi\rangle=|\phi\rangle_{A}\otimes|\Phi\rangle_{BC}$, with $|\Phi\rangle_{BC}$ entangled, gives an entangled reduced state,
$\rho^{(BC)}=|\Phi\rangle\langle\Phi|_{BC}.$ There are also non-pure states of this subtype: for instance, the family of
biseparable states
$\rho_{\epsilon}={1\over2}(1+\epsilon)|1\rangle\langle1|_{A}\otimes|\Psi_{+}\rangle\langle\Psi_{+}|_{BC}+{1\over2}(1-\epsilon)
|0\rangle\langle0|\otimes|\Psi_{-}\rangle\langle\Psi_{-}|_{BC}, |\epsilon|\leq1$, have reduced states
$\rho_{\epsilon}^{(BC)}={1\over2}(1+\epsilon)|\Psi_{+}\rangle\langle\Psi_{+}|_{BC}+{1\over2}(1-\epsilon)|\Psi_{-}\rangle
\langle\Psi_{-}|_{BC}$, with negativities $N(\rho_{\epsilon}^{(BC)})=|\epsilon|$ ; therefore these reduced states are entangled
for $\epsilon\neq0$ and the three qubit state $\rho_{\epsilon}$ is in subtype $1^1-1$; for $\epsilon=0$, $\rho_{{0}}$ is the
state previously considered as an example of subtype $1^1-0$; note that this $1^1-0$ state is a continuous limit
($\epsilon\rightarrow0$) of states of subtype $1^1-1$.
\item Subtype $1^2$: generalized biseparable states with bipartite entanglements in two pairs of qubits, for instance
$\rho = p_{1A}\ \rho_{1}^{A}\otimes\rho_{1}^{BC} + p_{1B}\ \rho_{1}^{B}\otimes\rho_{1}^{AC}+p_{2A}\
\rho_{2}^{A}\otimes\rho_{2}^{BC} + p_{2B}\ \rho_{2}^{B}\otimes\rho_{2}^{AC}$, with $\sum_{iJ}\ p_{iJ}=1$ and at least one
$\rho_{i}^{KL}$ entangled for each KL. According to the posible entanglements of their reduced states the following subtypes are
possible in principle: $1^2-0$ , $1^2-1$, $1^2-2$. We omit the graphic representation of these subtypes.
\item Subtype $1^3$: generalized biseparable states with bipartite entanglements in the three pairs of qubits, for instance:
$\rho=p_{A}\ \rho^{A}\otimes\rho^{BC}+p_{B}\ \rho^{B}\otimes\rho^{AC}+p_{C}\ \rho^{C}\otimes\rho^{AB}$, with $\sum_{J}\ p_{J}=1$
and the three $\rho^{KL}$ entangled. According to the posible entanglements of their reduced states the following subtypes are
possible in principle: $1^3-0$, $1^3-1$, $1^3-2$, $1^3-3$. We omit also the graphic representation of these subtypes.
Generalized biseparable states of subtypes $1^2$ and $1^3$ exist only in the non-pure case.
\item Type 2: fully inseparable states, non-zero full tripartite entanglement. Subtypes $2-0$ to $2-3$ exist for pure and non-pure states.
\item Subtype $2-0$: their three reduced entanglements are zero. We will call them \textit{GHZ-like states}, because the
well known GHZ states (see \ref{sec:4}) belong to this subtype. The entanglement of GHZ-like states disappears if any of the
three qubits is traced over; their entanglement is \textit{fragile}.
\item Subtype $2-1$: one reduced entanglement is non-zero, and the other two are zero.
\item Subtype $2-2$: two reduced entanglements are non-zero. For pure states they  have been called \textit{star shaped
states} \cite{pleuno}.
\item Subtype $2-3$: their three reduced entanglements are non-zero. We will call them \textit{W-like states} because the
well known $W$ states \cite{ciruno} (see also \ref{sec:4}) belong to this subtype. The entanglement in these states survives the
loss (tracing over) of any of the three qubits; it is \textit{robust}.
\end{itemize}
For pure states our classification is very easy to implement. A qubit is factorizable if and only if the reduced state of the
other two qubits is pure (see for instance chapter 8 of \cite{myfun}). In the affirmative case, this pure two-qubit state is
factorizable if and only if their reduced one-qubit states are pure; the total state is then fully separable (subtype $0-0$);
otherwise the reduced two-qubit state is entangled and the total state is biseparable (type $1^1-1$). If no qubit is
factorizable, the total state is fully entangled (type 2); the subtypes $2-0$, $2-1$, $2-2$, or $2-3$ can be ascertain
calculating the negativities or concurrences of their reduced two-qubit states to determine the number of entangled pairs. No
ambiguity remains in the qualitative application of our classification to pure states. In \ref{sec:4} we will show the results
for all possible GSD's of pure three-qubit states, obtaining the explicit form of the vectors in each one of our classes, and
the quantitative values of their entanglements.

For non-pure states the situation is more complicated. We do not know of any measure that would identify unambiguously
entanglement in general non-pure three-qubit states. Therefore, the discrimination between fully separable, biseparable and
fully entangled states remains open in general, although it can be answered in some particular cases, like those of the non-pure
examples of subtypes $1^1-0$ and $1^1-1$ given previously. In \ref{sec:3} we will propose a measure that, even if it is not a
complete solution, in some cases improves on previous results, as we shall see in the examples of \ref{sec:5}.

There are families of states depending on one or several parameters, such that by continuous variations of these the state goes
from one type to another. Or more generally, there are states $\Psi_1$,$\Psi_2$ ,  in two different types or subtypes such that
the distance $\|\Psi_1-\Psi_2\|$ is arbitrarily small; we will show examples in section 5.

Parts of this classification have antecedents in the literature. In \cite{pleuno}, a classification of pure three-qubit states
based only in the number of the entangled reduced states was proposed. But with only this criterion, fully separable states
(type $0-0$ in our classification) and GHZ-like states (type $2-0$) are in the same class; types $1^{i}-1$ (i=1,2,3) and $2-1$
are also in another common class, and so on. This classification leaves out the more important property of the entanglement of
three-qubit states: no entanglement, bipartite entanglements only, or full tripartite entanglement. In \cite{pledos}, the same
authors introduced new classes for non-pure states, taking into account the existence of classical correlations (separable but
non-factorizable states); in this paper we shall restrict ourselves to quantum correlations.

In \cite{cirdos} a classification of three-qubit entanglement in fully separable, biseparable (in the sense that we denote as
simply biseparable) or fully inseparable states was given; generalized biseparable states were not considered as a class
different from fully inseparable ones, and the existence or not of reduced binary entanglements was not considered.

\section{Some measures of full tripartite entanglement}
\label{sec:3}

An ideal measure of the full tripartite entanglement of three qubits should have at least the following characteristics: i) to
be zero for any fully separable or biseparable state and non-zero for any fully entangled state, ii) to be invariant under LU,
iii) to be non increasing under LOCC, that is, to be an entanglement monotone \cite{vid}.

Condition i) seems self evident, but some proposals for tripartite entanglement measures do not fulfill it, even for pure
states, as we will see below; the proposal that we will make (tripartite negativity) satisfies both parts of i) for pure states.
Conditions ii) and iii) are the mathematical expression of the non-local character of entanglement. Some authors include other
desirable conditions \cite{rad} but we will restrict ourselves to the three conditions listed above.

There are in the literature proposals for measures of tripartite entanglement of pure states, for instance those in \cite{coff},
\cite{chanuno},  \cite{mey},  \cite{bren},  \cite{pan}; besides specific objections that we will comment below, they cannot be
generally extended to non-pure states. In   \cite{chandos},   \cite{chantres} and   \cite{geo}, some particular results were
given for non-pure states.

In \cite{coff} a generalization of the bipartite concurrence, called the \textit{3-tangle}, is proposed. The original $W$ states
have 3-tangle equal to zero \cite{coff}; in fact, we found that the 3-tangle is zero for a large number of pure states in
Subtype $2-3$ ($W$-like), although none of their qubits is separable (the three reduced states $\rho^{(BC)}, \rho^{(AC)},
\rho^{(AB)}$ for these $W$-like states are all of them non-pure). Therefore, the 3-tangle is not a good measure of full
tripartite entanglement even for pure states; it is a measure of something that has been called \textit{residual entanglement}
by some authors \cite{loh}; in \cite{ciruno}  it was used to characterize one of the SLOCC classes (called $W-class$ by the
authors).

Yu and Song showed   \cite{chanuno} that any good measure $M_{A-B}$ of two-particle entanglement could be extended to
multiparticle systems, by taking bipartite partitions of them; they would define the following measure of tripartite
entanglement:
\begin{equation}
M^{add}_{ABC}={1\over 3}(M_{A-BC}+M_{B-AC}+M_{C-AB}).\label{eq:2}
\end{equation}
This is the idea underlying \cite{mey} and \cite{bren}, which generalize concurrence, and \cite{pan} which generalize Von
Neumman's entropy of reduced states.

Since the three terms in (2) verify separately conditions ii) and iii), so does also the tripartite additive measure
$M^{add}_{ABC}$. This was used in \cite{mey},  \cite{bren},  \cite{pan} and   \cite{lov} to prove that generalizations of
concurrence and Von Neumann's entropy of reduced states verify conditions ii) and iii).

But, with any of these elections, $M^{add}_{ABC}$ would be non-zero for pure biseparable states of Subtype $1^1-1$, violating
the first part of condition i). The same would also happen if a probability density function were used, as in \cite{pas}. It is
possible to avoid this objection by using the geometric mean instead of the arithmetic one:
\begin{equation}
M_{ABC}=(M_{A-BC}M_{B-AC}M_{C-AB})^{{1\over 3}} \label{eq:3}
\end{equation}
This idea was proposed, in a more general context, in \cite{lov}. So, if $M$ is the tangle (the square of the concurrence), we
have a multiplicative redefinition of the global entanglement $Q$ (whose additive version was introduced in \cite{mey} and
\cite{bren}), and if $M$ is Von Neumann's entropy of reduced states, we have a redefinition of the additive measure
$\eta^{(3)}_{\Psi}$ that appears in \cite{pan}. The same argument of the previous paragraph proves that these redefined product
versions of $Q$ and $\eta^{(3)}_{\Psi}$ verify conditions ii) and iii).

Von Neumann's entropy of reduced states is an unambiguous measure of entanglement only for pure states, and the concurrence,
although well defined for non-pure states of two qubits, has been extended in a practical way to higher dimensions only for pure
states. Therefore, in order to have a measure of tripartite entanglement valid also for non-pure states we will use the
negativity.

We will define the tripartite negativity of a state $\rho$ as
\begin{equation}
N_{ABC}(\rho)=(N_{A-BC}N_{B-AC}N_{C-AB})^{{1\over3}} \label{eq:4}
\end{equation}
where the bipartite negativities are defined as in (1),\\ $N_{I-JK}=-2\sum_{i}\sigma_{i}(\rho^{TI})$, $\sigma_{i}(\rho^{TI})$
being the negative eigenvalues of $\rho^{TI}$, the partial transpose of $\rho$ with respect to subsystem $I$,
$\langle{i_{I},j_{JK}}|{\rho^{TI}}|{k_{I},l_{JK}}\rangle=\langle{k_{I},j_{JK}}|{\rho}|{i_{I},l_{JK}}\rangle$, with $I=A, B, C$,
and $JK=BC, AC, AB$, respectively.

The bipartite negativities $N_{A-BC},N_{B-AC},N_{C-AB}$ verify conditions ii) and iii) \cite{vidwer}, and so our tripartite
negativity $N_{ABC}$ verifies them. For pure states, our multiplicative tripartite negativity fulfills the three conditions at
the beginning of this section. Although the qualitative classification of pure three-qubit states was easily done in
\ref{sec:2}, tripartite negativity adds a quantitative appraisal of the full tripartite entanglement of these states.

For non-pure two-party states in dimensions $2\times4$ there are entangled states with zero negativity \cite{horuno},
\cite{hordos}. Therefore, $N_{ABC}$ violates the second part of i). On the other hand, $N_{ABC}$ could be non-zero for
generalized biseparable states, violating also the first part of i). We could fulfill desideratum i) if we were able to replace
negativity for some other measure that will be non-zero for any entangled two-party state in dimension $2\times4$, and also
found a way to discriminate unambiguously between generalized biseparable and fully entangled states of three-qubits: these are
difficult and still open problems.

The set of three bipartite measures $N_{A-BC}$, $N_{B-AC}$, $N_{C-AB}$ contains more information than the geometric mean
$N_{ABC}$. But even this set cannot completely discriminate between fully separable, biseparable and fully entangled general
mixed states, and therefore does not improve essentially on the unique tripartite negativity $N_{ABC}$, at least for
classification purposes.

From the results in \cite{cirdos} it can be proved that $N_{ABC}>0$ is a sufficient condition for distillability to a GHZ state
(\textit{GHZ-distillability}), a property  of central importance in Quantum Computation \cite{akh}; therefore, tripartite
negativity is useful also for non-pure states, even if it does not solve the separability vs. entanglement problem.

\section{Explicit forms of pure states according to our classification}
\label{sec:4}

Our objective in this section is to divide the infinite set of LU equivalence classes of three-qubit pure states in six big
subsets corresponding to the six subtypes of Fig.1. Since a LU equivalence class is defined by a set of GSD coefficients, we
exhaustively analyzed the type of entanglement of all possible sets.

The full entanglement of the three-qubit state can be checked by studying its factorizability, as explained in \ref{sec:2}. The
entanglement of the reduced states (non-pure in general) will be determined calculating their negativity (for these mixed
two-qubit states, we could have used also the concurrence, but not von Neumann's entropy). The results are listed below.

In the following, we will denote the five coefficients of the GSD decomposition by ${\alpha,\beta,\delta, \epsilon,\omega}$; an
arbitrary three-qubit pure state can always be transformed to:
\begin{equation}
|\Psi\rangle=\alpha|{000}\rangle+\beta|{100}\rangle+\delta|{110}\rangle+\epsilon|{101}\rangle+\omega|{111}\rangle; \label{eq:5}
\end{equation}
this decomposition is symmetric in the interchange of the last two qubits, but not in the exchange of any of them with the
first. A more elegant totally symmetric GSD with five coefficients is also possible \cite{car}, but the algorithm to obtain the
coefficients from an arbitrary initial state is much more complicated; therefore it is not very useful for practical purposes.
Although the canonical form (5) contains five complex coefficients, it depends on only five independent real parameters; by
suitably choosing the relative phases for each qubit and the global phase of the state $|\Psi\rangle$ one can express it in
terms of five positive coefficients, related by the normalization condition, and a unique relevant phase, as it was done in
\cite{aciuno}. But starting from several generic vectors, different choices of phases will in general be needed to obtain a
simplified form of this kind; therefore, in order to compare the canonical forms for different initial vectors, we will postpone
any phase choices (and the normalization condition) until they are convenient; we will come back to this below when discussing
GHZ states.

Now we will show the canonical forms for our six entanglement subtypes:
\begin{itemize}
\item [i)]Type $0-0$: The conditions for fully separability are listed for instance in \cite{chancuatro}, in terms
of the eight coefficients of the Hilbert space. The GSD forms can be obtained from those conditions, taking into account that
three coefficients ($a_{001}, a_{010}, a_{011}$ in the notation of \cite{chancuatro}) are always zero.

\item [ii)]Subtype $1^1-1$: They have one of the following GSD forms:

$|{B}\rangle=\beta|{100}\rangle+\delta|{110}\rangle+\epsilon|{101}\rangle+\omega|{111}\rangle$ (where
$\beta\omega\neq\delta\epsilon$ and $\beta\omega$ or $\delta\epsilon$ can be zero; if $\beta\omega=\delta\epsilon$ the state is
of Type $0-0$).

$|{B^{'}}\rangle=\alpha|{000}\rangle+\beta|{100}\rangle+\epsilon|{101}\rangle$ (where $\beta$ can be zero)

$|{B^{''}}\rangle=\alpha|{000}\rangle+\beta|{100}\rangle+\delta|{110}\rangle$ (where $\beta$ can be zero).

For instance, for state $|{B}\rangle$ the bipartite reduced states $\rho^{(AB)}$, $\rho^{(AC)}$ are separable,  while
$\rho^{(BC)}$is entangled, with negativity $ N(\rho^{(BC)})=2|{\beta\omega-\delta\epsilon}|$.

\item [iii)] Subtype $2-0$ (GHZ-like states):

The states with these properties are of the canonical form $|{\Psi_{GHZ}}\rangle=\alpha|000\rangle+\omega|111\rangle$.

The three separable reduced states are of the form:
$\rho^{(BC)}=\rho^{(AC)}=\rho^{(AB)}=|\alpha|^{2}|00\rangle\langle00|+|\omega|^{2}|11\rangle
 \langle11|=|\alpha|^{2}|0\rangle\langle0|\otimes|0\rangle
 \langle0|+|\omega|^{2}|1\rangle\langle1|\otimes|1\rangle\langle1|$.

In particular, if the two coefficients are equal in modulus we will say that  we have a GHZ state, and denote it by
$|{GHZ}\rangle$. All the vectors of the form  ${1\over\sqrt2}(|000\rangle+e^{i\phi}|111\rangle)$, with $\phi$ real,  belong to
the bidimensional space of  GHZ states. With the election of canonical form of \cite{aciuno} only the final vector
${1\over\sqrt2}(|000\rangle+|111\rangle)$ would have been obtained; then, two orthogonal initial vectors in the GHZ class would
have been reduced to the same canonical vector,  while by postponing the choice of relative phases we have preserved the
dimensionality of the GHZ class and in particular the orthogonality relations in it.

Fig.2 shows several multiplicative measures of the full tripartite entanglement of GHZ-like states, as functions of the
coefficient $|\alpha|$ (now we impose the normalization condition $|\alpha|^{2} + |\omega|^{2}=1$).
 \begin{figure}[t]
 \begin{center}
 \includegraphics[width=0.9\textwidth]{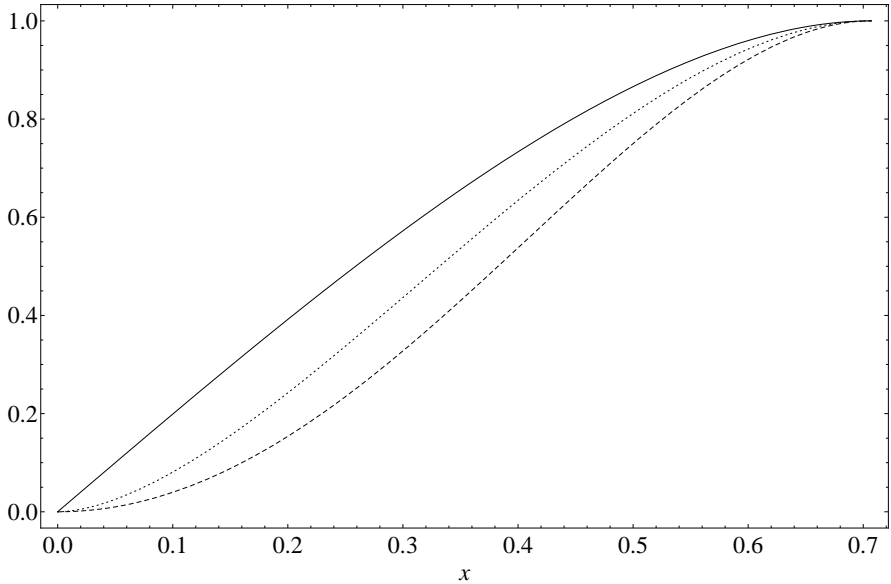}
 \caption{Comparison between $N_{ABC}$ (solid line), $Q$ (dash line) and $\eta^{(3)}_{\Psi}$ (dot line), as a function of $|\alpha|$
 for GHZ-like states.($|{\Psi_{GHZ}}\rangle=\alpha|000\rangle+\omega|111\rangle$, $|\alpha|\in[0,{1\over\sqrt2}]$; for $|\alpha|\in[{1\over\sqrt2},1]$ the graphic is symmetric and can be thought
off as representing  the simultaneous interchange of values 0 and 1 for the three qubits). The three measures induce the same
order for the full tripartite entanglement. GHZ states ($|\alpha|={1\over\sqrt{2}}$) are the ones with maximum full tripartite
entanglement ($N_{ABC}=Q=\eta^{(3)}_{\Psi}=1$). $N_{ABC}$ is more sensitive for small values of $|\alpha|$.}
 \end{center}
 \label{Fig:2}
 \end{figure}
The three measures induce the same order for the full tripartite entanglement; if a state has larger entanglement than other
with one measure, it has larger entanglement with the others. The negativity $N_{ABC}$ is more sensitive for small values of
$|\alpha|$.

We have performed a similar analysis for all the other classes, but the results depend on more than one parameter, and are more
difficult to represent: therefore we will not show them here. The results show in all cases that the three measures above
considered give qualitatively similar behaviours, and that the maximum full tripartite entanglement of any three-qubit state
corresponds to the LU equivalence class of GHZ states. Another explicit example will be given below when discussing Subtype
$2-3$.

\item[iv)] Subtype $2-1$: states of this class are those with one of the three following GSD forms:

$|{IV}\rangle=\alpha|000\rangle+\beta|100\rangle+\omega|111\rangle$,

$|{IV^{'}}\rangle=\alpha|000\rangle+\epsilon|101\rangle+\omega|111\rangle$,

$|{IV^{''}}\rangle=\alpha|000\rangle+\delta|110\rangle+\omega|111\rangle$,\\
with the three coefficients non-zero in each case.

For instance, for state $|{IV}\rangle$ the separable reduced states are $\rho^{(AB)}
=\rho^{(AC)}=(\alpha|0\rangle+\beta|1\rangle)(\alpha^{\star}\langle0|+\beta^{\star}\langle1|)\otimes
|0\rangle\langle0|+|\omega|^{2}|1\rangle\langle1|\otimes|1\rangle\langle1|$, while $\rho^{(BC)}$ is entangled, with negativity
$N(\rho^{(BC)})=2|\beta\omega|$. Note that if one of the three coefficients goes to zero, the state $|IV\rangle$ changes its
classification; in the limit $\beta=0$ we obtain a state of subtype $2-0$, in the limit $\alpha=0$ a state $|B\rangle$ of
subtype $1^1-1$, and in the limit $\omega=0$ a state of subtype $0-0$.

\item[v)] Subtype $2-2$ or star-shaped states: the GSD forms that have these properties are

$|{S}\rangle=\alpha|000\rangle+\beta|100\rangle+\epsilon|101\rangle+\omega|111\rangle$,

$|{S^{'}}\rangle=\alpha|000\rangle+\beta|100\rangle+\delta|110\rangle+\omega|111\rangle$,\\
with all the coefficients non-zero.

For instance, for state $|S\rangle$ the separable reduced state is $\rho^{(AB)}=(\alpha|0\rangle+\beta|1\rangle)
(\alpha^{\star}\langle0|+\beta^{\star}\langle1|)\otimes|0\rangle\langle0|+|1\rangle\langle1|\otimes(\epsilon|0\rangle+\omega|1\rangle)(\epsilon^{\star}\langle0|+\omega^{\star}
\langle1|)$, while $\rho^{(BC)}$ and $\rho^{(AC)}$ are entangled. In this case, although the negativities are perfectly
computable for any value of the coefficients, their analytic expressions in terms of generic coefficients are not very
comfortable. Thus, we give here the concurrences: $C(\rho^{(BC)})=2|\beta\omega|$, and $C(\rho^{(AC)})=2|\alpha\epsilon|$ (for
two-qubits, concurrence and negativity coincide for pure states, not for mixed states like $\rho^{(BC)}$, $\rho^{(AC)}$).

\item[vi)] Subtype $2-3$ or $W$-like class: states with the most general GSD form (five coefficients $\alpha$, $\beta$, $\epsilon$, $\delta$, $\omega$ different from zero)
belong to this class, and also those who have $\beta$ and/or $\omega$ equal to zero.

States with the canonical form $|\Psi_{W}\rangle=\alpha|000\rangle+\epsilon|101\rangle+\delta|110\rangle$, with the three
coefficients non-zero, are examples of this class. For these states, the bipartite reduced states are:

$\rho^{(BC)}=|\alpha|^{2}|00\rangle\langle00|+(\epsilon|01\rangle+\delta|10\rangle)(\epsilon^{\star}\langle01|+\delta^{\star}\langle10|)$,
$\rho^{(AC)}=(\alpha|00\rangle+\epsilon|11\rangle)(\alpha^{\star}\langle00|+\epsilon^{\star}\langle11|)+|\delta|^{2}|10\rangle\langle10|$,
and
$\rho^{(AB)}=(\alpha|00\rangle+\delta|11\rangle)(\alpha^{\star}\langle00|+\delta^{\star}\langle11|)+|\epsilon|^{2}|10\rangle\langle10|$,
with negativities: $N(\rho^{(BC)})=\sqrt{|\alpha|^{4}+4|\epsilon\delta|^{2}}-|\alpha|^{2}$;\\
$N(\rho^{(AC)})=\sqrt{|\delta|^{4}+4| \epsilon\alpha|^{2}}-|\delta|^{2}$;\\
$N(\rho^{(AB)})=\sqrt{|\epsilon|^{4}+4|\alpha\delta|^{2}}-|\epsilon|^{2}$.

In particular, if the three coefficients in $|\Psi_{W}\rangle$ are equal in modulus we will denote the state by $|W\rangle$ and
refer to it as a $W$ state; the original symmetric $W$ state \cite{ciruno},
\begin{equation}
|W^{'}\rangle={1\over\sqrt3}(|001\rangle+|010\rangle+|100\rangle). \label{eq:6}
\end{equation}
has a GSD of this form.

The maximum values of the tripartite entanglement measures in the $W$-like class, are reached for the $|W\rangle$ states:
$N_{ABC}=0.94$, $Q={8\over9}$, $\eta^{(3)}_{\Psi}=0.92$.

The negativities of the reduced bipartite states of the states $|W\rangle$ (which are the same as the negativities for the
symmetric $|W^{'}\rangle$ state) are $N(\rho^{(BC)})=N(\rho^{(AC)})=N(\rho^{(AB)})= {\sqrt{5}-1\over3}=0.41$.
\end{itemize}

Our classification of pure three-qubit states in terms of full tripartite entanglement and reduced binary entanglements could be
used to give a physical interpretation to the abstract GSD classes of \cite{aciuno} and \cite{acidos}.

\section{Discussion of some mixed three-qubit states}
\label{sec:5}

We do not have a canonical form like the GSD of the pure case to simplify the general form of mixed states; therefore we will
restrict ourselves to consider some concrete uniparametric families of non-pure states, to show some of the improvements allowed
by the use of tripartite negativity, the problems remaining and relations with previous works.

As a first example we will consider a family of mixed GHZ and $W$ states: $\rho(p)=p|GHZ\rangle\langle
GHZ|+(1-p)|W^{'}\rangle\langle W^{'}|$ where $|W^{'}\rangle$ has been defined in (6) and $0\leq{p}\leq1$. Its tripartite
negativity is \\$N_{ABC}(\rho)=N_{A-BC}={\sqrt{41p^{2}-64p+32}+2\sqrt{10p^{2}-2p+1}-p-2\over6}$
\\($N_{A-BC}=N_{B-AC}=N_{C-AB}$),\ \ that is different from zero for any value of the parameter $p$ (see Fig. 3), and has an absolute maximum
for pure GHZ states and a secondary maximum for pure $W$ states (Fig.3).

This excludes full separability or simple biseparability for any state in the family, but can not discriminate between
distributed binary entanglement and full tripartite entanglement.  Nevertheless this result improves on \cite{loh}, where the
3-tangle for this family was found to be zero for a value of the parameter $p$ (generalized biseparability was not considered as
distinct
from full entanglement in \cite{loh}). \\
\begin{figure}[t]
\begin{center}
  \includegraphics[width=0.9\textwidth]{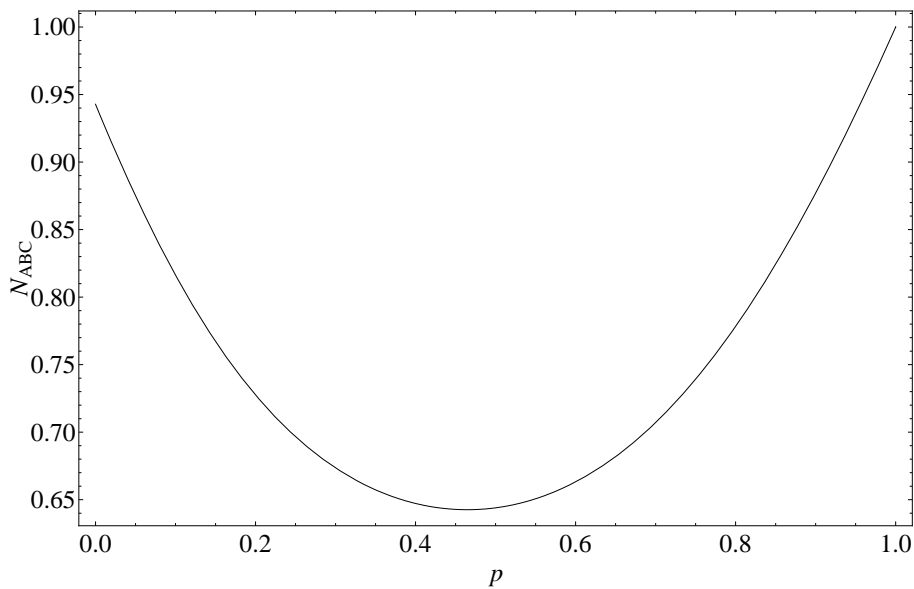}
\caption{Tripartite negativity for mixed GHZ and $W^{'}$ states, $\rho(p)=p|GHZ\rangle\langle GHZ|+(1-p)|W^{'}\rangle\langle
W^{'}|$, as a function of p.}
\end{center}
\end{figure}

Our second example is a family of mixed states with non-zero tripartite negativity $N_{ABC}$ only for some values of a
parameter: $\rho(p)=p|GHZ\rangle\langle{GHZ}|+{1-p\over8}\textbf1$, where \textbf{1} is the unity matrix, and $0\leq p\leq1$. We
find that $N_{ABC}(\rho)=0$ if $p\leq{1\over5}$, and $N_{ABC}(\rho)={5p-1\over4}$ if $p>{1\over5}$, reaching a maximum
$N_{ABC}(\rho)=1$ for $p=1$ (a pure GHZ state).

D\"ur et al. \cite{cirdos} showed  that this state is GHZ-distillable if $p>{1\over5}$. Therefore, for this family of mixed
states our tripartite negativity quantifies GHZ-distillability; it starts at $p={1\over5}$ and increases with $p$, with an
obvious maximum of 1 if the state is a GHZ state.\\

Finally, we will consider a family of three-qubit states that can be fully entangled or generalized biseparable and nevertheless
have zero tripartite negativity. In \cite{hordos}, a family of mixed $2\times4$ states of the form\bigskip\\
$\sigma_{b}={1\over{7b+1}}\pmatrix{b&0&0&0&0&b&0&0\cr 0&b&0&0&0&0&b&0\cr 0&0&b&0&0&0&0&b\cr 0&0&0&b&0&0&0&0\cr
0&0&0&0&{1+b\over2}&0&0&{\sqrt{1-b^{2}}
\over2}\cr b&0&0&0&0&b&0&0\cr 0&b&0&0&0&0&b&0\cr 0&0&b&0&{\sqrt{1-b^{2}}\over2}&0&0&{1+b\over2}}$\bigskip\\
with $0< b <1$ was considered. These states have zero negativity for any value of the parameter b, although they were proved to
be non-separable. We do not know of any practical measure of bipartite entanglement that will discriminate these bound entangled
states from separable ones. Biconcurrence \cite{bad}, could do this in principle; unfortunately its determination needs the
calculus of the minimum of a certain function for any unitary operator in a Hilbert space of dimension 64, and no efficient way
to do this is known \cite{rew}.

We can convert these states in dimension $2\times4$ to three-qubit states simply by taking the following basis in the 4
dimensional space:\newline $\
\{|e_{1}\rangle,|e_{2}\rangle,|e_{3}\rangle,|e_{4}\rangle\}=\{|00\rangle_{BC},|01\rangle_{BC},|10\rangle_{BC},|11\rangle_{BC}\}$.

The bipartite negativities are then  $N_{B-AC}=N_{C-AB}={\sqrt{3b^2+1}-2b\over7b+1}$ $>0$, $N_{A-BC}=0$ $\forall b$, $0<b<1$,
and therefore $N_{ABC}=0$. According to \cite{hordos}, the states are not separable into A and BC subsystems. Since the other
two negativities are non-zero, we also know that they are not fully separable nor simply biseparable. Thus, this family of
states can be fully entangled or generalized biseparable.

\section{Conclusions}
\label{sec:6}

We have proposed a classification of three-qubit states based on the existence of bipartite and tripartite entanglements and the
diverse possibilities for the reduced binary entanglements, including a graphic representation for pure states that can be
extended to non-pure ones, although we have not done it here for reasons of space. We have considered a measure (tripartite
negativity) of the full tripartite entanglement avoiding some of the problems of previous proposals; for pure states this
measure quantifies full tripartite entanglement and confirms the distinction between fully entangled states and biseparable or
fully separable ones that was obtained in \ref{sec:2} in a qualitative way. We have given also the explicit form of the pure
states in each subtype of our classification, after performing an easily computable GSD to a simplified canonical form; in the
simplest cases we have compared their tripartite negativity with other multiplicative generalizations of bipartite entanglement
measures, concluding that they induce the same ordering of full tripartite entanglement. We have analyzed some non-pure states
that have non-zero tripartite negativity (a sufficient condition for GHZ-distillability) or that have zero tripartite negativity
although they are known to be entangled, to show the problems that remain in the practical classification of mixed three-qubit
states.\\

We thank G. \'Alvarez, D. Salgado, L. Lamata and J. Le\'on for valuable discussion and help. C.S. acknowledge financial support
from CSIC I3 program.


\begin{thebibliography}{99}
\bibitem{gur}L. Gurvits, Journal of Computer and System Sciences \textbf{69}, 448 (2004).
\bibitem{aciuno}A. Ac\'in, A. Andrianov, L.Costa, E. Jan\'e, J.I. Latorre, R. Tarrach, Phys. Rev. Lett. \textbf{85}, 7 (2000).
\bibitem{acidos}A. Ac\'in, A. Andrianov, E. Jan\'e, R. Tarrach, J. Phys. A: Math. Gen. \textbf{34}, 6725 (2001).
\bibitem{ciruno}W. D\"ur, G. Vidal and J. Cirac, Phys. Rev. A \textbf{62}, 062314 (2000).
\bibitem{luc} L. Lamata, J. Le\'on, D. Salgado and E. Solano,  Phys. Rev. A 74, 052336 (2006).
\bibitem{pleuno}M. Plesch, V. Bu\v zek, Phys. Rev. A \textbf{67}, 012322 (2003).
\bibitem{pledos}M. Plesch, V. Bu\v zek, Phys. Rev. A \textbf{68}, 012313 (2003).
\bibitem{coff}V. Coffman, J.Kundu, W.K. Wootters, Phys. Rev A \textbf{61}, 0532306 (2000).
\bibitem{chanuno}Chang-shui Yu, He-shan Song, Phys. Lett A 330, 377 (2004).
\bibitem{mey}D. Meyer, N.R. Wallach, J. of Math. Phys., \textbf{43}, pp.4273 (2002).
\bibitem{bren}G.K. Brennen, Quantum Information and Computation, vol. 3 (6), 619-626 (2003).
\bibitem{pan}F. Pan, D. Liu, G. Lu, J.P. Draayer, Int.J.Theor.Phys. \textbf{43}, 1241 (2004).
\bibitem{pas}P.Facchi, G. Florio and S. Pascazio, Phys. Rev. A \textbf{74}, 042331 (2006).
\bibitem{wer}R.F. Werner, Phys. Rev. A \textbf{40},4277 (1989).
\bibitem{rad}T. Radtke, S. Fritzsche, Comput. Phys. Comm. \textbf{175}, 145 (2006).
\bibitem{sak}J.J. Sakurai, \textit{Modern Quantum Mechanics}, Addison-Wesley
Publishing Company (1994), 183-184.
\bibitem{woo}W.K. Wootters, S. Hill, Phys. Rev. Lett \textbf{78}, 5022-5025 (1997) .
\bibitem{vidwer}G. Vidal, R.F. Werner, Phys. Rev. A \textbf{65}, 032314 (2002).
\bibitem{bar}S. Barn, S.J.D. Phoenix, Phys. Rev. A \textbf{44}, 535 (1991).
\bibitem{rung}P. Rungta, V. Buzek, C. M. Caves, M. Hillery, G. J. Millburn, Phys. Rev. A, \textbf{64}, 042315 (2001).
\bibitem{rew}R. Horodecki, M. Horodecki, P. Horodecki, K. Horodecki, quant-ph/070225.
\bibitem{bad} P.Badziag, P. Deuar, M. Horodecki, P. Horodecki, and R. Horodecki, J. Mod. Opt. \textbf{49}, 1289 (2002).
\bibitem{mir}A. Miranowicz, A. Grudka, J. Opt B: Quantum Semiclass. Optics 6 542-548 (2004) .
\bibitem{mirdos}A. Miranowicz, A. Grudka, Phys. Rev. A \textbf{70}, 032326 (2004).
\bibitem{plen}M.B. Plenio, S. Virmani, Quant. Inf. Comp. \textbf{7}, 1 (2007).
\bibitem{plendos}M.B. Plenio, Phys. Rev. Lett. \textbf{95}, 090503 (2005).
\bibitem{horuno}M. Horodecki, P. Horodecki, R. Horodecki, Phys. Rev. Lett. \textbf{80} 5239-5242 (1998) .
\bibitem{hordos}P. Horodecki, Phys. Lett. A, \textbf{232}, 333 (1997).
\bibitem{acitres}A. Ac\'in, D. Bru\ss, M.Lewnstein, A. Sanpera, Phys. Rev. Lett. \textbf{87}, 040401 (2001).
\bibitem{dru}D.Bru\ss \ et al. Phys. Rev. A \textbf{72}, 014301 (2005).
\bibitem{cirdos}W. D\"ur, J.I. Cirac, R. Tarrach, Phys. Rev. Lett. \textbf{83} 3562-3565 (1999).
\bibitem{egg}T. Eggeling and R. F. Werner, Phys. Rev. A \textbf{63}, 042111 (2001).
\bibitem{lask}W. Laskowki, M. Zukowski, Phys. Rev. A, \textbf{72}, 062112 (2005).
\bibitem{myfun}L.E. Ballentine, \textit{Quantum mechanics: a modern development}, World Scientific Publishing, 1998.
\bibitem{vid}G. Vidal, J. Mod. Opt. \textbf{47}355 (2000).
\bibitem{chandos}Chang-shui Yu, He-Shan Song, Phys. Rev. A \textbf{73}, 022325 (2006).
\bibitem{chantres}Chang-shui Yu, He-Shan Song, Phys. Rev. A \textbf{73}, 032322 (2006).
\bibitem{geo}Tzu-chieh Wei, Paul M. Goldbart, Phys. Rev. A \textbf{68}, 024307 (2003).
\bibitem{loh}R. Lohmayer, A. Osterloh, J. Siewert, A. Uhlmann, Phys. Rev. Lett. \textbf{97}, 260502 (2006).
\bibitem{lov}P.J. Love et al., Quantum Information Processing \textbf{6}, 187 (2007).
\bibitem{akh} S.J. Akhtarshenas, J. Phys. A \textbf{38}, 6777 (2005).
\bibitem{car}H. A. Carteret, A. Higuchi, A. Sudbery, J. Math. Phys. \textbf{41}, (2000) 7932-7939.
\bibitem{chancuatro}Chang-shui Yu, He-Shan Song, Phys. Rev. A \textbf{72}, 022333 (2005).

\end{thebibliography}
\end{document}